\documentclass[a4paper,12pt]{article}

\usepackage[margin=2.5cm]{geometry}
\usepackage{graphicx}
\usepackage{amssymb}
\usepackage{multirow}
\usepackage{amsmath}
\usepackage{setspace}
\usepackage{caption}
\usepackage{subcaption}
\usepackage{cases}
\usepackage[colorlinks=true,bookmarks=true]{hyperref}
\usepackage{float}
\usepackage{mathrsfs}
\usepackage[nosort]{cite}

\hypersetup{citecolor=blue}
\newcommand{\dif}{\mathrm{d}}

\makeatletter
\renewcommand\section{\@startsection {section}{1}{\z@}%
	{-3.5ex \@plus -1ex \@minus -.2ex}%
	{2.3ex \@plus.2ex}%
	{\normalfont\bfseries}}
\renewcommand\subsection{\@startsection{subsection}{2}{\z@}%
	{-3.25ex\@plus -1ex \@minus -.2ex}%
	{1.5ex \@plus .2ex}%
	{\normalfont\it}}% from \large
\renewcommand\subsubsection{\@startsection{subsubsection}{3}{\z@}%
	{-3.25ex\@plus -1ex \@minus -.2ex}%
	{1.5ex \@plus .2ex}%
	{\normalfont}}% from \normalsize
\makeatother

\begin{document}

\thispagestyle{empty}
\begin{flushright}
  
\end{flushright}
\vbox{}
\vspace{2cm}

\begin{center}
  {\LARGE{Black holes with bottle-shaped horizons%\\[2mm]
  }}\\[16mm]
  {{Yu Chen~~and~~Edward Teo}}
  \\[6mm]
    {\it Department of Physics,
      National University of Singapore, %\\[1mm]
      Singapore 119260}\\[15mm]
\end{center}
\vspace{2cm}

\centerline{\bf Abstract}
\bigskip
\noindent
We present a new class of four-dimensional AdS black holes with non-compact event horizons of finite area. The event horizons are topologically spheres with one puncture, with the puncture pushed to infinity in the form of a cusp. Because of the shape of their event horizons, we call such black holes ``black bottles''. 
The solution was obtained as a special case of the Pleba\'nski--Demia\'nski solution, and may describe either static or rotating black bottles. 
For certain ranges of parameters, an acceleration horizon may also appear in the space-time. We study the full parameter space of the solution, and the various limiting cases that arise. In particular, we show how the rotating black hole recently discovered by Klemm arises as a special limit.

\newpage

\tableofcontents

\section{Introduction}		

There are well-known theorems which state that black holes in four-dimensional, asymptotically flat space-times must have spherical event-horizon topology \cite{Hawking:1971vc,Friedman:1993ty}. Attempts to directly extend these theorems to more general situations, however, have largely been unsuccessful. Indeed, it was realised quite early on that black holes in an asymptotically anti-de Sitter (AdS) space-time can have the topology of a Riemann surface with arbitrary genus \cite{Lemos:1994xp,Lemos:1994fn,Lemos:1995cm,Cai:1996eg,Mann:1996gj,Vanzo:1997gw,Brill:1997mf,Mann:1997iz,Birmingham:1998nr}. Such solutions are known as ``topological black holes''. More recently, asymptotically flat black holes with ring-shaped horizons have been discovered in higher dimensions (see, e.g., \cite{Emparan:2008eg} for a review). 

The class of topological black holes discovered in \cite{Lemos:1994xp,Lemos:1994fn,Lemos:1995cm,Cai:1996eg,Mann:1996gj,Vanzo:1997gw,Brill:1997mf,Mann:1997iz,Birmingham:1998nr} is described by the static metric
\begin{align}
\label{top_bh}
\dif s^2&=-f(r)\dif t^2+f(r)^{-1}\dif r^2+r^2\dif\Sigma_{(k)}^2\,,\cr
f(r)&=k-\frac{2m}{r}+\frac{r^2}{\ell^2}\,,
\end{align}
where $\ell$ is related to the cosmological constant $\Lambda$ by $\ell^2=-\frac{3}{\Lambda}$, and $m$ is the mass parameter of the black hole. $\dif\Sigma_{(k)}^2$ is a 2-surface with constant scalar curvature $2k$. We may normalise $k$ so that it takes the values $k=\pm1$ and 0, in which case $\dif\Sigma_{(k)}^2$ has the form
\begin{subequations}
\begin{numcases}
{\dif\Sigma_{(k)}^2=}
\dif\rho^2+\sin^2\rho\,\dif\phi^2, & $k=+1$; \\
\dif\rho^2+\dif\phi^2, & $k=0$; \\
\dif\rho^2+\sinh^2\rho\,\dif\phi^2, & $k=-1$. \label{hyperbolic_plane}
\end{numcases}
\end{subequations}
Note that for constant $t$, the horizon geometry is a sphere, plane or hyperbolic plane for $k=+1,0,-1$ respectively. In the latter two cases, the surfaces described by $\dif\Sigma_{(k)}^2$ are non-compact, but they can be made compact by appropriate identifications on the coordinates. For the $k=0$ case, the topology of the horizon can be turned into that of a torus, while for the $k=-1$ case, it can be turned into that of a Riemann surface with genus greater than one.

\begin{figure}
	\begin{center}
		\includegraphics[width=3in,angle=-90]{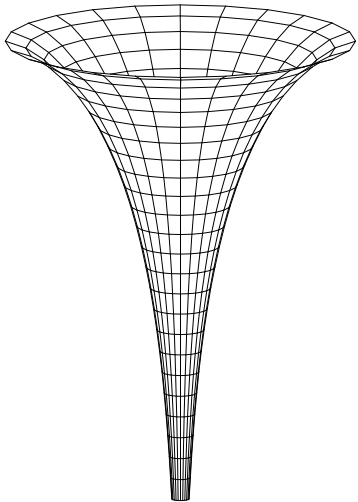}
		\caption{The cusp, as embedded as a surface of revolution in a three-dimensional Euclidean space. The top of the surface is where $\rho=0$, while $\rho\rightarrow\infty$ is reached asymptotically at the bottom of the figure.}
		\label{fig_cusp}
	\end{center}
\end{figure}

In this paper, we will be exclusively focussed on the case in which the horizon is a hyperbolic surface. Moreover, for reasons that will become clear below, we will only be interested in non-compact hyperbolic surfaces. Besides the hyperbolic plane metric $\dif\Sigma_{(-1)}^2$ in (\ref{hyperbolic_plane}), there exist two other non-compact constant-curvature hyperbolic surfaces whose metrics: 
\begin{subequations}
\begin{align}
\dif\Sigma_{(-1)}^{'2}&=\dif\rho^2+\cosh^2\rho\,\dif\phi^2,\\
\dif\Sigma_{(-1)}^{''2}&=\dif\rho^2+{\rm e}^{-2\rho}\dif\phi^2,
\label{cusp}
\end{align}
\end{subequations}
can be substituted into (\ref{top_bh}). These two surfaces can be obtained from the hyperbolic plane by quotienting it by an appropriate subgroup of its isometry group (see, e.g., \cite{Borthwick}). For reasons explained in \cite{Borthwick}, they are known as the hyperbolic and parabolic cylinders respectively. While the hyperbolic cylinder has two ends $\rho\rightarrow\pm\infty$ that are symmetric under the reflection $\rho\rightarrow-\rho$, the parabolic cylinder has non-symmetric ends. In particular, the small end of the parabolic cylinder with $\rho>0$ is known as a ``cusp'', and can be embedded in a three-dimensional Euclidean space as shown in Fig.~\ref{fig_cusp}.\footnote{It should be pointed out that it is not possible in general to embed a hyperbolic surface in a three-dimensional Euclidean space. The cusp is an exception to this rule.}  The cusp has the remarkable property that it has a finite area.

An early attempt to generalise the hyperbolic black hole (\ref{top_bh}) and (\ref{hyperbolic_plane}) was to include a rotation parameter \cite{Klemm:1997ea}. However, it was realised (see the errata of \cite{Klemm:1997ea}) that when rotation is present, the coordinate identifications that turn the hyperbolic surface into a Riemann surface of higher genus {\it cannot\/} be performed consistently. The horizon of the rotating hyperbolic black hole has to remain non-compact. Asymptotically, where the effects of the rotation vanish, it approaches the form of (\ref{hyperbolic_plane}). In this sense, it is a {\it deformed\/} hyperbolic surface, with the rotation parameter determining the amount of deformation away from (\ref{hyperbolic_plane}).

The rotating hyperbolic black hole, like its spherical counterpart, belongs to the Carter--Pleba\'nski solution \cite{Carter:1968ks,Plebanski}. Now, the Carter--Pleba\'nski solution is a special case of the well-known Pleba\'nski--Demia\'nski solution \cite{Plebanski:1976gy}, which contains an extra parameter commonly known as an acceleration parameter. Very recently, a static class of hyperbolic black holes with a non-zero acceleration parameter was derived from this solution \cite{Chen:2015zoa}. It was shown that they have horizons which are asymptotically a hyperbolic surface, similar to that of the rotating hyperbolic black hole. Thus the acceleration parameter, like the rotation parameter, determines the amount of deformation of the horizon away from (\ref{hyperbolic_plane}).

\begin{figure}
	\begin{center}
		\includegraphics[width=3in,angle=-90]{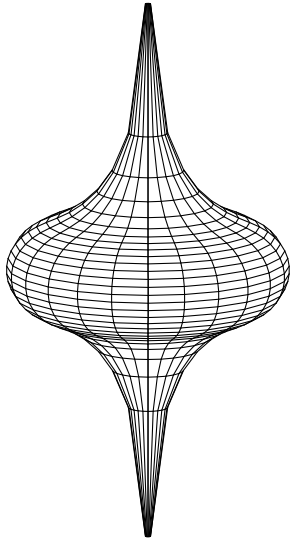}
		\caption{The black spindle, as embedded as a surface of revolution in a three-dimensional Euclidean space. It has cusps extending to infinity at the top and bottom of the figure.}
        \label{fig_spindle}
	\end{center}
\end{figure}

At this stage, one might wonder if there exist black holes whose horizons are asymptotically cusps of the form (\ref{cusp}), instead of the form (\ref{hyperbolic_plane}). Such a horizon would have a finite area, in contrast to that of the rotating or accelerating hyperbolic black hole which has an infinite area. Indeed, such a solution was recently discovered by Klemm \cite{Klemm:2014rda} (see also \cite{Gnecchi:2013mja}). This solution describes a rotating black hole whose horizon has two cusp ends, as shown in Fig.~\ref{fig_spindle}. Because of its shape, we shall refer to this black hole as a ``black spindle''. It was found as a special case of the Carter--Pleba\'nski solution, and can also be obtained as an ultra-spinning limit of the Kerr--AdS solution. In particular, the latter implies that the black spindle does not have a static limit. Further properties of this solution have been studied in \cite{Hennigar:2014cfa,Hennigar:2015cja}.

In hindsight, it is clear why the black spindle is necessarily rotating. We have seen that a static hyperbolic black hole must have a horizon with a constant negative scalar curvature. However, the horizon of the black spindle in Fig.~\ref{fig_spindle} obviously does not have this property. In fact, it has a central spherical region with a {\it positive\/} scalar curvature. It is only along the two cusps that the horizon becomes hyperbolic, so the horizon is what we refer to as a deformed hyperbolic surface. Such a horizon is compatible with the presence of a non-trivial rotation parameter.

Topologically, the horizon of the black spindle is a sphere with two punctures. The space-time is made complete by pushing these punctures to infinity, in the form of cusp ends. Thus despite the non-compact nature of the horizon, its area remains finite. This construction has an obvious extension to an arbitrary number of punctures. However, if we restrict ourselves to black-hole space-times with an axial symmetry, then the only other possibility is a horizon that is topologically a sphere with {\it one\/} puncture. Such a horizon would look like a one-ended version of Fig.~\ref{fig_spindle} (c.f.\ Fig.~\ref{fig_bottle} below), and we shall call such a black hole a ``black bottle''.

Such a black bottle solution was described in \cite{Chen:2015zoa}, as a special case of the class of solutions considered in that paper. It arises from the static limit of the Pleba\'nski--Demia\'nski solution, so it has a vanishing rotation parameter but a non-vanishing acceleration parameter. It is this acceleration parameter which allows the horizon to be asymptotically hyperbolic, with a single cusp end. This static black bottle actually predates the rotating black spindle discovered by Klemm: it is the so-called planar black hole in the black droplet solution of \cite{Hubeny:2009ru}.

In this paper, we would like to consider generalising the black bottle to include rotation. It turns out that such a rotating black bottle does exist within the general Pleba\'nski--Demia\'nski solution, although finding a form of it which is amenable to analysis is not immediately obvious. To this end, we use some of the ideas developed in \cite{Chen:2015vma,Chen:2015zoa} for the AdS C-metric, to find a suitable form for both the static and rotating black bottles. In the static case, it actually differs from the form used in \cite{Chen:2015zoa}. The advantage of this new form becomes apparent when the rotating case is considered; for example, the parameter space describing rotating black bottles is relatively simple and can be completely characterised in this form.

The organisation of this paper is as follows: We begin in Sec.~\ref{static_bb} with an analysis of the static black bottle. We present the new form of this solution, find the appropriate coordinate and parameter ranges, and study its geometrical and physical properties. In Sec.~\ref{rotating_bb}, we turn our attention to the rotating black bottle and build upon the analysis of the static case. In particular, we present the full two-dimensional parameter space of this solution. We also study the various special cases of this solution, including how the black spindle arises as a limiting case of the black bottle. The paper concludes with a summary and discussion of the results. There is also an appendix in which we show how the black bottle solution can be derived from the Pleba\'nski--Demia\'nski solution.

\section{Static black bottle}
\label{static_bb}

The metric describing a static black bottle is given by
\begin{align}
\label{metric_static_droplet}
\dif s^2&=\frac{\ell^2(1-b)}{(x-y)^2}\bigg[Q\dif t^2-\frac{\dif y^2}{Q}+\frac{\dif x^2}{P}+P\dif \phi^2\bigg]\,,\cr
P&=1+x-x^2-x^3,\qquad Q=b+y-y^2-y^3.
\end{align}
We note that the functions $P$ and $Q$ can also be written as
\begin{align}
\label{structure_functions_static}
P(x)=(1-x)(1+x)^2,\qquad Q(y)=P(y)-(1-b)\,.
\end{align}
This solution has two parameters: $\ell$ and $b$. The former is related to the cosmological constant $\Lambda$ by 
\begin{align}
\label{Lambda}
\ell^2=-\frac{3}{\Lambda}\,,
\end{align}
and sets the scale of the space-time. Since we are only interested in AdS space-times, we can assume $\ell^2>0$. The solution (\ref{metric_static_droplet}) was identified in \cite{Chen:2015zoa} as a subclass of the AdS C-metric. In Appendix \ref{sec_appendix}, we explicitly show how it can be derived from the general Pleba\'nski--Demia\'nski solution.

\subsection{Coordinate and parameter ranges}

We now wish to find the appropriate coordinate and parameter ranges, such that (\ref{metric_static_droplet}) describes a space-time free of curvature singularities and with the correct Lorentzian signature ($-$+++). It can be checked that there are in general curvature singularities located at $x,y=\pm\infty$, so that the coordinates $x$ and $y$ must take finite ranges. Furthermore, they must satisfy either $x<y$ or $x>y$, since $x=y$ is where conformal infinity of the space-time (\ref{metric_static_droplet}) is located. The requirement of Lorentzian signature means that either
\begin{subequations}
\begin{align}
\label{b<1}
P>0,\quad Q<0 \quad\hbox{and}\quad b<1\,,
\end{align}
or
\begin{align}
\label{b>1}
P<0,\quad Q>0 \quad\hbox{and}\quad b>1\,.
\end{align}
\end{subequations}
It follows that the ranges of $x$ and $y$ will be bounded by the roots of $P$ and $Q$ respectively, in addition to the line $x=y$.

The roots of $P$ can be trivially read off from the first equation in (\ref{structure_functions_static}): there is a single root at $x=+1$ and a double root at $x=-1$. It follows that the appropriate finite range of $x$ is $-1<x<+1$, for which $P>0$. This rules out the case (\ref{b>1}), leaving just (\ref{b<1}) to consider.

On the other hand, the root structure of $Q$ depends on the value of the parameter $b$. In the range $b<1$, $Q$ may possess either one or three real roots. It turns out that the case in which $Q$ possesses only one real root describes a black bottle without an acceleration horizon, while the case in which $Q$ possesses three real roots describes a black bottle with an acceleration horizon. We now study these two cases separately.

\subsubsection{Static black bottle without an acceleration horizon}

\begin{figure}
 \begin{center}
  \begin{subfigure}[b]{0.45\textwidth}
   \centering
   \includegraphics[scale=0.45]{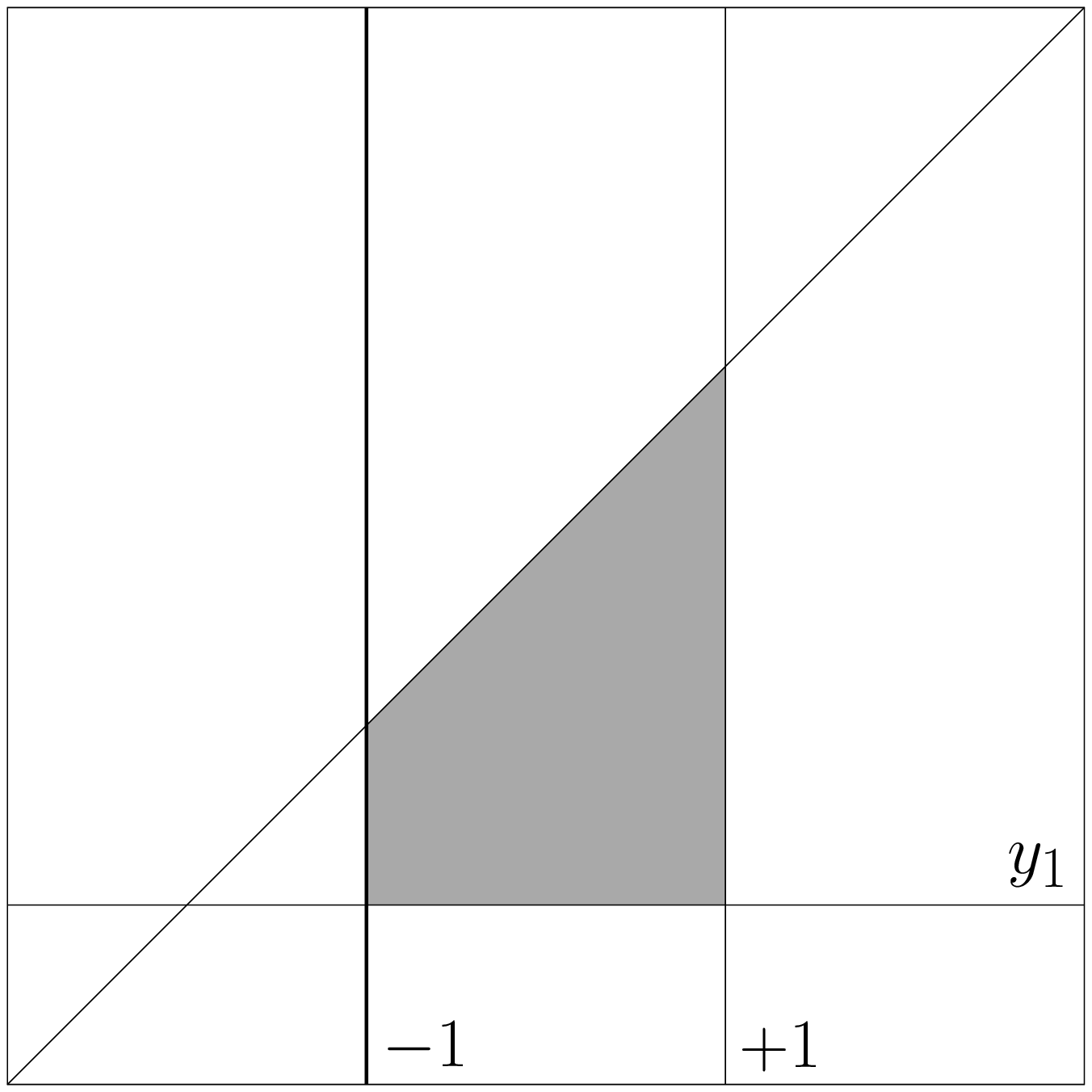}
   \caption{}
   \label{fig_domain_static_bottle}
  \end{subfigure}
  \begin{subfigure}[b]{0.45\textwidth}
   \centering
   \includegraphics[scale=0.45]{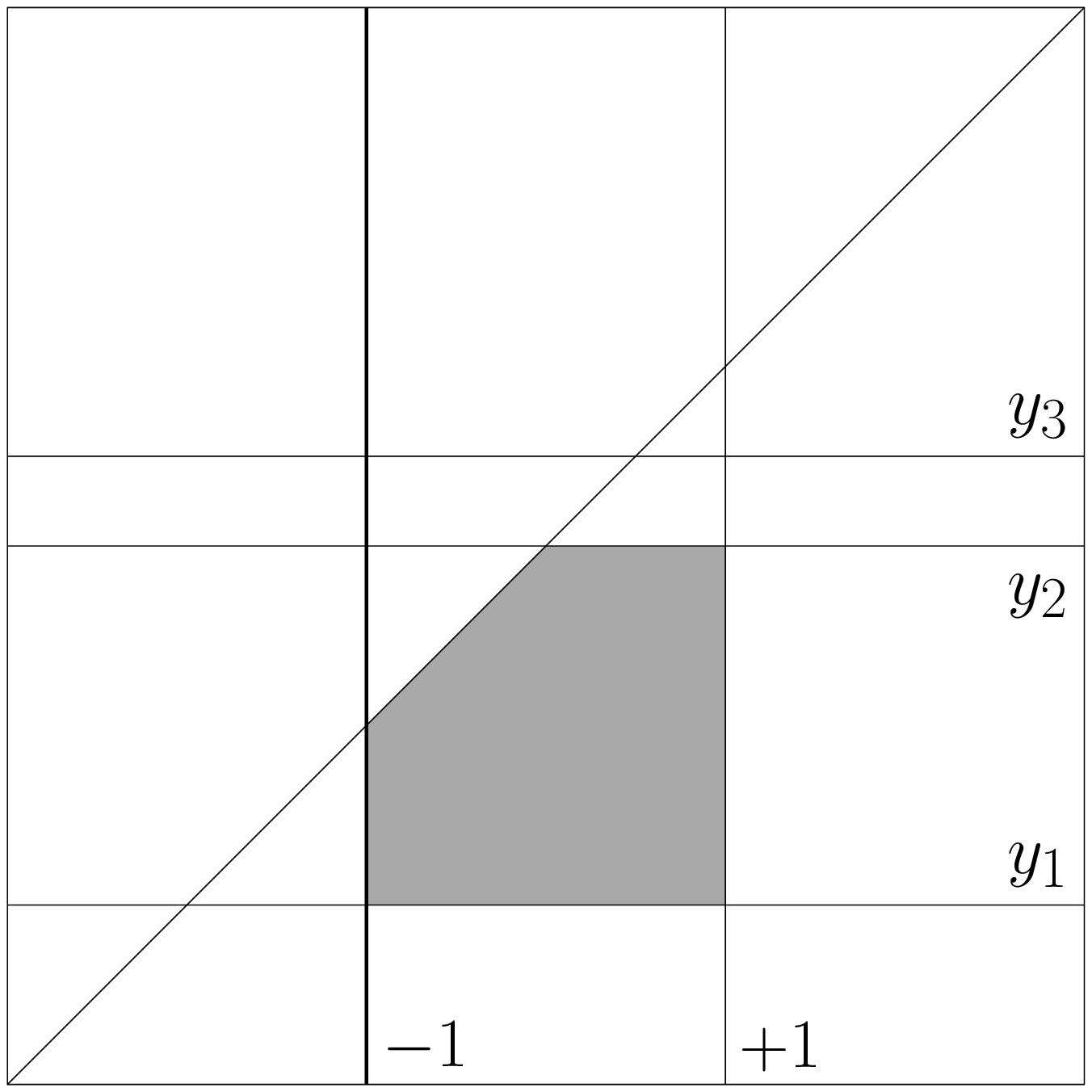}
   \caption{}
   \label{fig_domain_static_droplet}
  \end{subfigure}
 \end{center}
 \caption{The domains of the (a) static black bottle without an acceleration horizon, and (b) static black bottle with an acceleration horizon. The $x$ coordinate runs in the horizontal direction, while the $y$ coordinate runs in the vertical direction. The shaded areas depict the coordinate ranges of interest. The diagonal line corresponds to $x=y$, while the edges of the plot correspond to $x,y=\pm\infty$.}
\end{figure}

We first consider the range
\begin{align}
b<-\frac{5}{27}\,,
\end{align}
in which case $Q$ has only one real root $y_1$, satisfying
\begin{align}
y_1<-\frac{5}{3}<-1\,.
\end{align}
It follows that if we restrict ourselves to the following ranges of $x$ and $y$:
\begin{align}
\label{domain_static_bottle}
-1<x<+1\,,\qquad y_1<y<x\,,
\end{align}
the metric (\ref{metric_static_droplet}) will describe a space-time with Lorentzian signature. We refer to the coordinate range (\ref{domain_static_bottle}) as the domain of the space-time, and it is illustrated in Fig.~\ref{fig_domain_static_bottle}.

The physical meaning of the boundaries of the domain will be discussed in Sec.~\ref{sec_static_properties}. We just mention here that $y=y_1$ corresponds to a Killing horizon in the space-time. It is in fact a black-hole horizon, since there is a curvature singularity at $y=-\infty$ beneath it. As we shall see below, the double root at $x=-1$ endows the black-hole horizon with a bottle shape, so we shall call this black hole a ``black bottle''. Unlike the following case we consider, there are no other horizons present in the space-time.

\subsubsection{Static black bottle with an acceleration horizon}

In the remaining range to consider:
\begin{align}
-\frac{5}{27}\leq b<1\,,
\label{range_static_droplet}
\end{align}
$Q$ has three real roots. If we denote them by $y_{1,2,3}$ in increasing order, they satisfy
\begin{align}
-\frac{5}{3}\leq y_1<-1<y_2\leq y_3<+1\,.
\end{align}
It follows that if we restrict ourselves to the following ranges of $x$ and $y$:
\begin{align}
\label{domain_static_droplet}
-1<x<+1\,,\qquad y_1<y<y_2\,,\qquad y<x\,,
\end{align}
the metric (\ref{metric_static_droplet}) will describe a space-time with Lorentzian signature. The domain of this space-time is illustrated in Fig.~\ref{fig_domain_static_droplet}.

As in the previous case, $y=y_1$ corresponds to a Killing horizon that is a black-hole horizon. It again has the shape of a bottle. In this case, however, there is a second Killing horizon at $y=y_2$, which forms the upper boundary of the domain. This Killing horizon extends to conformal infinity $x=y$, and can be understood to be an acceleration horizon of the space-time \cite{Podolsky:2002nk,Dias:2002mi,Krtous:2005ej}. This means that the black bottle is undergoing an acceleration, with $b$ as the acceleration parameter. In the previous case, the acceleration of the black bottle is sufficiently small that the acceleration horizon is absent. It is in fact the analogue of the ``slowly accelerating'' spherical black holes that are known to exist in AdS space-time \cite{Podolsky:2002nk,Dias:2002mi,Krtous:2005ej}.

Because of its shape, the Killing horizon at $y=y_2$ has also been referred to as a ``black droplet'' by Hubeny et al.~\cite{Hubeny:2009ru}.\footnote{To make contact with the metric studied in \cite{Hubeny:2009ru}, we start from (\ref{metric_static_droplet}) and perform the following coordinate transformations and parameter redefinition:
\begin{align}
	x\rightarrow \frac{3+4\sqrt{3}x}{9}\,,\qquad y\rightarrow \frac{3+4\sqrt{3}y}{9}\,,\qquad (\psi,\phi)\rightarrow \frac{3\sqrt{3}}{8} (\psi,\phi)\,,\qquad b=-\frac{5+32\lambda}{27}\,.\nonumber
\end{align}
We then obtain \cite{Hubeny:2009ru}
\begin{align}
	\dif s^2&=\frac{\ell^2(1+\lambda)}{(x-y)^2}\bigg[{F}\dif t^2-\frac{\dif y^2}{{F}}+\frac{\dif x^2}{{G}}+{G}\dif \phi^2\bigg]\,,\nonumber\\
	F&=-\lambda-y^2-2\mu y^3,\qquad G=1-x^2-2\mu x^3,\nonumber
\end{align}
with $\mu=1/(3\sqrt{3})$. The parameter range (\ref{range_static_droplet}) corresponds to $-1<\lambda\leq0$, in which case $F$ has three real roots.} We will investigate the shape of this horizon in more detail in Sec.~\ref{horiz_geom}.

We remark that the critical case $b=-\frac{5}{27}$ (or equivalently $y_1=-\frac{5}{3}$) occurs when $y_2=y_3$. This corresponds to the case in which the acceleration horizon has become extremal. When the acceleration horizon is pushed beyond extremality, it disappears from the system. This occurs when $b<-\frac{5}{27}$, which corresponds to the black bottle without an acceleration horizon considered previously.

On the other hand, the limit $b\rightarrow1$ is well defined only if $\ell$ is sent to infinity while keeping $\ell^2(1-b)$ finite, resulting in a Ricci-flat metric. As will be explained below, this limit is of no interest to us.

\subsection{Geometrical and physical properties}
\label{sec_static_properties}

\subsubsection{Rod structure}

Having obtained the possible domains of the black bottle solution, we now turn to a study of the boundary of each domain. It turns out that this will give us much useful information about the geometrical and physical properties of the space-time.

We have already mentioned that the boundary segment $x=y$ represents conformal infinity of the space-time. The remaining segments of the boundary represent sets of space-time points at which either $P=0$ or $Q=0$. They can also be understood to be sets of points at which some linear combination of the Killing vector fields $\frac{\partial}{\partial t}$ and $\frac{\partial}{\partial\phi}$:
\begin{align}
k\propto \alpha\frac{\partial}{\partial t}+\beta\frac{\partial}{\partial\phi}\,,
\end{align}
has vanishing norm. If $k$ is time-like, the segment represents a Killing horizon of the space-time; if $k$ is space-like, the segment represents an axis of the space-time. Each such segment is known as a ``rod'', and can be defined to have the normalised direction:
\begin{align}
k&=\frac{1}{\kappa}\left(\alpha\frac{\partial}{\partial t}+\beta\frac{\partial}{\partial\phi}\right)\cr
&\equiv\frac{1}{\kappa}(\alpha,\beta)\,.
\end{align}
Here, $\kappa$ is the surface gravity of the Killing horizon, if $k$ is time-like and $\alpha$ is chosen to have unit value. If $k$ is space-like, $\kappa$ is a Euclidean version of the surface gravity \cite{Chen:2010zu}. The collection of all the rods, together with their directions, is known as the rod structure of the space-time. More details of the rod-structure formalism can be found in \cite{Chen:2010zu}, and references therein.

We first consider the rod structure of the black bottle with an acceleration horizon; that of the black bottle without an acceleration horizon will arise as a special case of this. It turns out to have the following rod structure:
\begin{itemize}
	\item Rod 1: a semi-infinite space-like rod located at $(x\,{=}\,{-1},y_1\,{\leq}\,y\,{<}\,{-1})$, with direction\footnote{Since $\kappa_{\text{E}1}=0$ in this case, $k_1$ is strictly speaking not normalisable. However, we continue to adopt the formal notation for $k_1$ in (\ref{k1}), since it usefully encodes value of the (Euclidean) surface gravity. The implications of the vanishing of $\kappa_{\text{E}1}$ will be discussed below.}
	\begin{align}
        \label{k1}
	k_1=\frac{1}{\kappa_{\text{E}1}}(0,1)\,,\qquad \kappa_{\text{E}1}=0\,;
	\end{align}
	
	\item Rod 2: a finite time-like rod located at $(-1\,{\le}\,x\,{\le}\,{+1},y\,{=}\,y_1)$, with direction
	\begin{align}
	k_2=\frac{1}{\kappa_2}(1,0)\,,\qquad \kappa_2=-\frac{1}{2}\frac{\dif Q}{\dif y}\bigg|_{y=y_1}\,;
	\end{align}
	
	\item Rod 3: a finite space-like rod located at $(x\,{=}\,{+1},y_1\,{\le}\,y\,{\le}\,y_2)$, with direction
	\begin{align}
	k_3=\frac{1}{\kappa_{\text{E}3}}(0,1)\,,\qquad \kappa_{\text{E}3}=2\,;
	\end{align}
	
	\item Rod 4: a semi-infinite time-like rod located at $(y_2\,{<}\,x\,{\le}\,{+1},y\,{=}\,y_2)$, with direction
	\begin{align}
	k_4=\frac{1}{\kappa_4}(1,0)\,,\qquad \kappa_4=\frac{1}{2}\frac{\dif Q}{\dif y}\bigg|_{y=y_2}\,.
	\end{align}
\end{itemize}
These four rods meet at the three turning points $(x\,{=}\,{-1},y\,{=}\,y_1)$, $(x\,{=}\,{+1},y\,{=}\,y_1)$ and $(x\,{=}\,{+1},y\,{=}\,y_2)$. Note that Rods 1 and 3 correspond to the left and right vertical boundaries of the domain in Fig.~\ref{fig_domain_static_droplet} respectively, while Rods 2 and 4 correspond to the lower and upper horizontal boundaries respectively. Rods 1 and 4 are semi-infinite, since the boundaries they correspond to are joined up to conformal infinity $x=y$.

Since Rods 2 and 4 are time-like, they are Killing horizons in the space-time. As we will see in the following subsection, the horizon described by Rod 2 has the shape of a bottle; it is the black bottle. On the other hand, the horizon described by Rod 4 extends to conformal infinity; it is an acceleration horizon. The surface gravities of these two horizons are $\kappa_2$ and $\kappa_4$ respectively, and they are both guaranteed to be non-negative by the profile of the function $Q$.

Turning to the space-like rods, we note that Rod 3 is an axis that stretches between the two horizons. The Euclidean surface gravity of this rod, $\kappa_{\text{E}3}$, actually encodes the natural periodicity of the relevant azimuthal coordinate, in this case the $\phi$ coordinate. To avoid a conical singularity along this axis, the coordinate identification
\begin{align}
\label{identification_static}
(t,\phi)\rightarrow \left(t,\phi+\frac{2\pi}{\kappa_{{\rm E}3}}\right)=(t,\phi+\pi)\,,
\end{align}
has to be made.

On the surface, Rod 1 would seem to be another axis of the space-time. However, note that it has vanishing Euclidean surface gravity $\kappa_{\text{E}1}$. The reason for this can be traced to the fact that $x=-1$ is a double root of the function $P$, and it actually implies that Rod 1 is at an infinite proper distance from the other points in the space-time. This can be seen from the fact that the integral
\begin{align}
\int_{-1}^{x_0}\frac{1}{x-y}\frac{\dif x}{\sqrt{P}}=\int_{-1}^{x_0}\frac{\dif x}{(x-y)\sqrt{1-x}\,(x+1)}\,,
\end{align}
diverges. Thus, Rod 1 represents a new spatial infinity of the space-time, distinct from conformal infinity $x=y$.

There is in fact a similar result for time-like rods. If two roots of $Q$ were to coincide, they would describe an extremal horizon with vanishing surface gravity. As is well known from say the Reissner--Nordstr\"om or Kerr black hole, such an extremal horizon is infinitely far away from the other points in the space-time.

We now briefly describe the rod structure of the black bottle without an acceleration horizon. It can in fact be obtained from the above rod structure by simply removing the fourth rod. Without this rod, Rod 3 will extend to conformal infinity $x=y$; it now has coordinates $(x=+1,y_1\le y<+1)$. The three rods of this rod structure clearly form the vertical and horizontal boundaries of the domain in Fig.~\ref{fig_domain_static_bottle}.

As in the case considered above, Rod 2 is a Killing horizon; it is the so-called black bottle. Rod 3 is an axis of the space-time that now extends from the horizon to conformal infinity. To avoid a conical singularity along this axis, the same coordinate identification as in (\ref{identification_static}) has to be made. Lastly, Rod 1 forms a new spatial infinity of the space-time. 

In both cases, note that the bottle horizon described by Rod 2 extends to the new spatial infinity described by Rod 1. This is the first hint that the so-called neck of the bottle is infinitely long. Moreover, the $\phi$-circle vanishes at the end of the neck. We will perform a detailed analysis of this horizon geometry in the following subsection.

\subsubsection{Horizon geometries}
\label{horiz_geom}

To study the geometry of the black-hole horizon represented by Rod 2, it is convenient to reparameterise the solution in terms of $y_1$. This amounts to writing $b$ in terms of $y_1$ as\footnote{\label{footnote3}In this parameterisation, we note that $y_{2,3}$ can be explicitly expressed as
\begin{align}
y_{2,3}=-\frac{1+y_1\pm \sqrt{(1-y_1)(5+3y_1)}}{2}\,.\nonumber
\end{align}}
\begin{align}
b=y_1(-1+y_1+y_1^2)\,.
\end{align}
For a constant time slice, the horizon has the induced metric
\begin{align}
\label{horizon_metric}
\dif s_{\text{BH}}^2=\frac{\ell^2(1-y_1)(1+y_1)^2}{(x-y_1)^2}\bigg[\frac{\dif x^2}{{(1-x)(1+x)^2}}+{(1-x)(1+x)^2}\dif \phi^2\bigg]\,.
\end{align}
It can be checked that the coordinate identification (\ref{identification_static}) ensures that the ``north pole'' of the horizon, $x=+1$, is a regular point of the space-time. The geometry around this point resembles that around the north pole of a perfect sphere.

The ``south pole'' of the horizon, $x=-1$, however, presents a different story. If we define the new coordinate $\rho\equiv-\ln(1+x)$, the south pole can be approached by taking the limit $\rho\rightarrow\infty$. In this limit, the metric (\ref{horizon_metric}) approaches
\begin{align}
\label{south_pole_static}
\dif s_{\text{BH}}^2\rightarrow\frac{\ell^2(1-y_1)}{2}\big(\dif \rho^2+4{\rm e}^{-2\rho}\dif \phi^2\big)\,,
\end{align}
which is the cusp metric (\ref{cusp}), up to a conformal factor and a rescaling of $\phi$. As can be seen from (\ref{south_pole_static}), the south pole is infinitely far away from the other points of the geometry. Moreover, the size of the $\phi$-circle goes to zero as the south pole is approached.

\begin{figure}
	\begin{center}
		\includegraphics[width=3in,angle=-90]{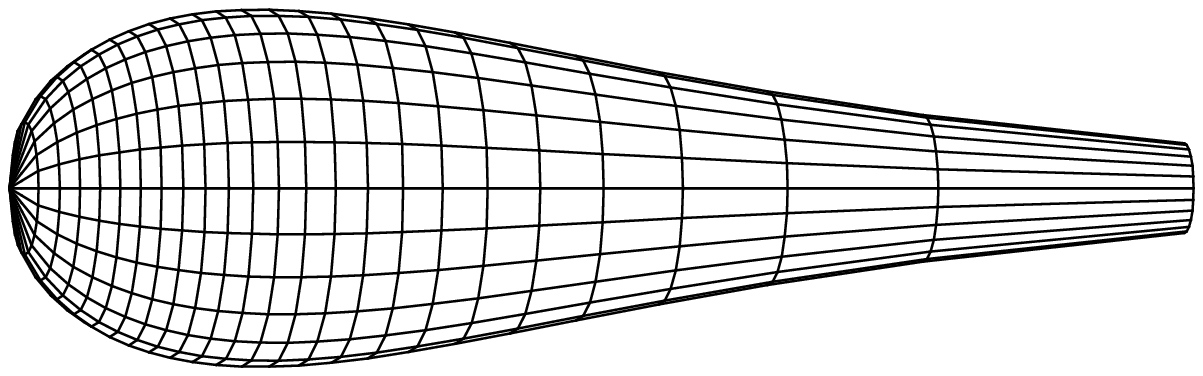}
		\caption{The black bottle, as embedded as a surface of revolution in a three-dimensional Euclidean space. The north pole of the bottle is connected by an axis to conformal infinity at the top of the figure. The neck of the bottle extends to the new spatial infinity at the bottom of the figure.}
		\label{fig_bottle}
	\end{center}
\end{figure}

A fuller picture of the horizon geometry can be obtained from the scalar curvature of the metric (\ref{horizon_metric}), which can be calculated to be
\begin{align}
R=-\frac{2}{\ell^2}+\frac{2(x-y_1)^3}{\ell^2(1-y_1)(1+y_1)^2}\,.
\end{align}
At the north and south poles, it is
\begin{align}
R(x\,{=}\,{+1})=-\frac{8y_1}{\ell^2(1+y_1)^2}>0\,,\qquad R(x\,{=}\,{-1})=\frac{4}{\ell^2(y_1-1)}<0\,.
\end{align}
Note that the scalar curvature is positive around the north pole, indicating a spherical geometry there. On the other hand, the scalar curvature is negative around the south pole, indicating a hyperbolic geometry there. Thus, the horizon interpolates between a spherical geometry and a hyperbolic one. 

This picture is confirmed if we embed (\ref{horizon_metric}) as a surface of revolution in a three-dimensional Euclidean space. This embedding is illustrated in Fig.~\ref{fig_bottle}. As can be seen, the geometry near the north pole is spherical, while the geometry near the south pole is that of a cusp. Because its shape resembles an (inverted) bottle, we shall call this black hole a ``black bottle''. The cusp forms the infinitely long and thin neck of the bottle.

We note that the size of the $\phi$-circle shrinks exponentially to zero at the south pole. This behaviour leads to the finiteness of the area of the horizon, which can be straightforwardly calculated to be
\begin{align}
A_{\text{BH}}=-2\pi \ell^2(1+y_1)\,.
\end{align}
We thus see that $y_1$ has an interpretation as the area parameter of the bottle horizon. In the limit $b\rightarrow -\infty$ for fixed $\ell$, the area of the horizon blows up. In the limit $b\rightarrow 1$ with $\ell^2(1-b)$ fixed, the area also blows up. So a well-behaved black bottle solution does not exist in either limit. In particular, the latter limit indicates that the black bottle does not exist in Ricci-flat space-times.

\begin{figure}
	\begin{center}
		\includegraphics[width=3in,angle=-90]{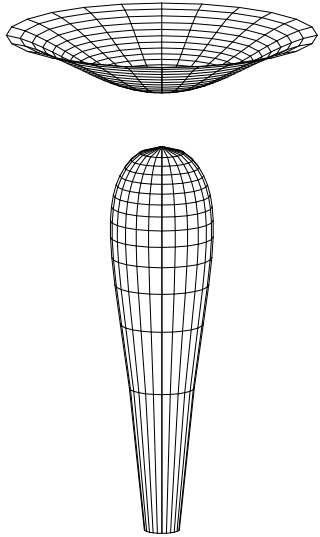}
		\caption{The black bottle with an acceleration horizon, as embedded as a surface of revolution in a three-dimensional Euclidean space. The bottle connects to the acceleration horizon by an axis, and the latter extends to conformal infinity at the top of the figure. As usual, the neck of the bottle extends to the new spatial infinity at the bottom of the figure.}
		\label{fig_droplet}
	\end{center}
\end{figure}

Turning to the acceleration horizon represented by Rod 4, we can similarly calculate its induced metric:
\begin{align}
\label{horizon_metric2}
\dif s_{\text{AH}}^2=\frac{\ell^2(1-y_2)(1+y_2)^2}{(x-y_2)^2}\bigg[\frac{\dif x^2}{{(1-x)(1+x)^2}}+{(1-x)(1+x)^2}\dif \phi^2\bigg]\,,
\end{align}
where $y_2$ is given in terms of $y_1$ by the equation in Footnote \ref{footnote3}. It can be checked that the coordinate identification (\ref{identification_static}) ensures that the ``north pole'' of the horizon, $x=+1$, is a regular point of the space-time. At the other end of the range, $x=y_2$, the conformal factor in front of the metric (\ref{horizon_metric2}) blows up. Thus the horizon extends to conformal infinity of the space-time, and has an infinite area.

The scalar curvature of the metric (\ref{horizon_metric2}) can be calculated to be
\begin{align}
R=-\frac{2}{\ell^2}+\frac{2(x-y_2)^3}{\ell^2(1-y_2)(1+y_2)^2}\,.
\end{align}
At the north pole and at conformal infinity, it is
\begin{align}
R(x\,{=}\,{+1})=-\frac{8y_2}{\ell^2(1+y_2)^2}\,,\qquad 
R(x\,{=}\,y_2)=-\frac{2}{\ell^2}<0\,.
\end{align}
Note that the scalar curvature is always negative at $x=y_2$. However, the scalar curvature can take either sign at $x=+1$: it is positive if $y_2$ is negative, and {\it vice versa\/}. 

When $R(x\,{=}\,{+1})$ is positive, the geometry is spherical around the north pole. The horizon thus interpolates between a spherical geometry at the north pole, and a hyperbolic one at conformal infinity. In this case, it is possible to partially embed (\ref{horizon_metric2}) as a surface of revolution in a three-dimensional Euclidean space. Such an embedding is illustrated in Fig.~\ref{fig_droplet}, superimposed on an embedding of the corresponding bottle horizon. The embedding fails at some stage before the horizon reaches conformal infinity at the top of the figure. This horizon has the shape of a droplet hanging from above; that is why it is also known as a ``black droplet'' \cite{Hubeny:2009ru}.

When $R(x\,{=}\,{+1})$ is negative, the geometry is hyperbolic around the north pole. In this case, it is not possible to embed (\ref{horizon_metric2}) in a three-dimensional Euclidean space. Nonetheless, the horizon will still have the same topology as in the previous case, and it might be appropriate to continue referring to it as a black droplet.

\section{Rotating black bottle}
\label{rotating_bb}

The rotating generalisation of the metric (\ref{metric_static_droplet}) is given by
\begin{align}
\dif s^2&=\frac{\ell^2(1-b)}{(x-y)^2}\bigg[\frac{{Q}}{1+a x^2y^2}(\dif t-\sqrt{a}x^2\dif \phi)^2-\frac{1+a x^2y^2}{{Q}}\,\dif y^2\nonumber\\
&\hspace{0.75in}+\frac{1+a x^2y^2}{{P}}\,\dif x^2+\frac{{P}}{1+a x^2y^2}(\dif \phi+\sqrt{a}y^2\dif t)^2\bigg]\,,\nonumber\\
{P}&=1+(1+ab)x-(1-ab)x^2-(1+ab)x^3-abx^4,\nonumber\\
{Q}&=b+(1+ab)y-(1-ab)y^2-(1+ab)y^3-ay^4.
\label{metric_rotating_droplet}
\end{align}
We note that the functions $P$ and $Q$ can also be written as
\begin{align}
\label{structure_functions_rotating}
P(x)=(1-x)(1+x)^2(1+abx)\,,\qquad Q(y)=P(y)-(1-b)(1+ay^4)\,.
\end{align}
This solution has three parameters: $\ell$, $a$ and $b$. The static solution (\ref{metric_static_droplet}) is recovered by setting $a=0$, so $a$ can be interpreted as a rotation parameter. The other two parameters $\ell$ and $b$ have the same interpretations as in the static case. This solution can be derived from the general Pleba\'nski--Demia\'nski solution, as we explicitly show in Appendix~\ref{sec_appendix}.

\subsection{Coordinate and parameter ranges}

We now wish to find the appropriate coordinate and parameter ranges, such that (\ref{metric_rotating_droplet}) describes a space-time free of curvature singularities and with the correct Lorentzian signature. Moreover, these ranges should reduce to those found for the static case in the limit $a=0$.

Now, it can be checked that there are in general curvature singularities located at $(x\,{=}\,{\pm\infty},y\,{=}\,0)$ and $(x\,{=}\,0,y\,{=}\,{\pm\infty})$, so the desired coordinate range must avoid these points. Recall that in the static case, the requirement of Lorentzian signature means that (\ref{b<1}) must hold. Moreover, in this case, we need to impose $a\geq0$, so that the metric (\ref{metric_rotating_droplet}) remains real. Thus, to ensure the correct signature, we impose the ranges of parameters:
\begin{align}
a\ge 0\,,\qquad b<1\,.
\label{range_ab1}
\end{align}
Again, the Ricci-flat limit $b\rightarrow1$ is of no interest to us.

The roots of $P$ can be read off from the first equation in (\ref{structure_functions_rotating}). As in the static case, there is a single root at $x=+1$ and a double root at $x=-1$. But there is now a new root at $x=-\frac{1}{ab}$. The appropriate range of $x$ is still between the original two roots, i.e., $-1<x<+1$. The requirement that $P$ is positive in this range then gives the constraint
\begin{align}
-1\le ab\le 1\,.
\label{range_ab2}
\end{align}

This range of $x$ means that the curvature singularities at $(x\,{=}\,{\pm\infty},y\,{=}\,0)$ will be avoided. To avoid the other curvature singularities at $(x\,{=}\,0,y\,{=}\,{\pm\infty})$, we follow the static case and demand that $Q$ has at least one real root $y_1$, satisfying $y_1<-1$, which forms the lower bound for the range of $y$. The upper bound for $y$ is taken to be the next root of $Q$ greater than $y_1$, if it exists. In any case, an absolute finite upper bound for $y$ is $y=x$, which represents conformal infinity of the space-time (\ref{metric_rotating_droplet}). In this range of $y$, we require that $Q$ is negative for Lorentzian signature.

Since $Q$ is a quartic polynomial with negative leading coefficient (the case $a=0$ will no longer be considered), we in fact have the stronger result that it admits at least two real roots $y_{0,1}$, satisfying
\begin{align}
y_0\leq y_1<-1\,.
\label{y0y1}
\end{align}
If $Q$ admits four real roots, then the other two roots will satisfy
\begin{align}
-1<y_2\leq y_3<+1\,.
\end{align}
This follows from the simple inequalities:
\begin{align}
Q(-1)&=Q(+1)=-(1+a)(1-b)<0\,,\nonumber\\
Q'(-1)&=4a(1-b)>0\,,\qquad Q'(+1)=-4(1+a)<0\,.
\end{align}
It is of course also possible that $y_2$ and $y_3$ are both complex, so that $Q$ has only two real roots at $y_0$ and $y_1$. It turns out that the case in which $Q$ possesses only two real roots describes a rotating black bottle without an acceleration horizon, while the case in which $Q$ possesses four real roots describes a rotating black bottle with an acceleration horizon. We now study these two cases separately.

\subsubsection{Rotating black bottle without an acceleration horizon}

This case corresponds to $Q$ having only two real roots, at $y_0$ and $y_1$. The domain of interest is then
\begin{align}
\label{domain1}
-1<x<+1\,,\qquad y_1<y<x\,.
\end{align}
This domain structure is similar to that of the static black bottle without an acceleration horizon in Fig.~\ref{fig_domain_static_bottle}. In particular, the bottle horizon continues to be located at $y=y_1$. The differences are that there is an extra root of $P$ at $x=-\frac{1}{ab}$ (which may lie either on the left or right of the domain), and an extra root of $Q$ at $y=y_0\leq y_1$. The latter can in fact be interpreted as the inner horizon of the rotating black bottle. When $y_0$ and $y_1$ become degenerate, we have an extremal rotating black bottle without an acceleration horizon.

\subsubsection{Rotating black bottle with an acceleration horizon}

In the case, $Q$ has four real roots satisfying
\begin{align}
y_0\leq y_1<-1<y_2\leq y_3<+1\,.
\end{align}
The domain of interest is then
\begin{align}
\label{domain2}
-1<x<+1\,,\qquad y_1<y<y_2\,,\qquad y<x\,.
\end{align}
Again, this domain structure is similar to that of the static black bottle with an acceleration horizon in Fig.~\ref{fig_domain_static_droplet}. In particular, the bottle horizon continues to be located at $y=y_1$, and the acceleration horizon at $y=y_2$. The differences are that there is an extra root of $P$ at $x=-\frac{1}{ab}$, and an extra root of $Q$ at $y=y_0\leq y_1$. The root at $y=y_0$ can again be interpreted as the inner horizon of the rotating black bottle. When $y_0$ and $y_1$ become degenerate, we have an extremal rotating black bottle with an acceleration horizon.

\subsubsection{Parameter space}

\begin{figure}
	\begin{center}
		\includegraphics[scale=1.3]{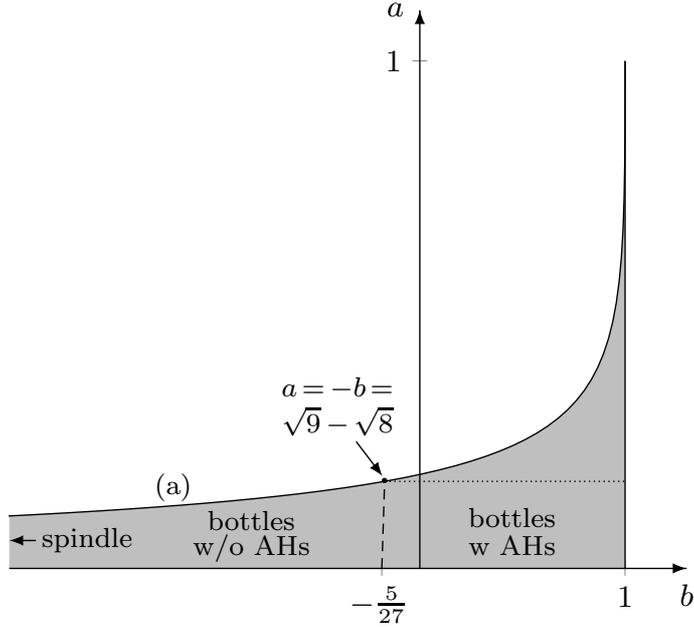}
		\caption{The parameter space describing rotating black bottles. The curve labelled by (a) corresponds to extremal black bottles, the dashed curve to black bottles with extremal acceleration horizons, and the dotted curve to black bottles in thermal equilibrium with their acceleration horizons. The dashed curve separates the black bottles with acceleration horizons (AHs) from those without. The black spindle is recovered in the limit when $b\rightarrow-\infty$ appropriately.}
		\label{fig_parameter_space}
	\end{center}
\end{figure}

It is of interest to understand the full $a$-$b$ parameter space describing rotating black bottles. Recall that we had imposed the constraints (\ref{range_ab1}) and (\ref{range_ab2}) on $a$ and $b$. It turns out that a much stronger constraint on the parameters comes from (\ref{y0y1}). As we shall see, the full parameter space, as an $a$-$b$ plot, is given by the shaded region in Fig.~\ref{fig_parameter_space}. It is bounded by three curves: the extremal black bottle curve labelled by (a), the static limit $a=0$, and the Ricci-flat limit $b=1$. This space is divided into subregions representing different physical configurations, the details of which will be explained in Sec.~\ref{sec_limits}. At this stage, we simply note the following global ranges of parameters implied by Fig.~\ref{fig_parameter_space}:
\begin{align}
0\leq a< 1\,,\qquad b<1\,.
\label{ab}
\end{align}
It can also be checked that the shaded region of Fig.~\ref{fig_parameter_space} lies within the range (\ref{range_ab2}), as it should. In fact, it touches the curves $|ab|=1$ only at $b=1$, and in the limit $b\rightarrow-\infty$. Since $b=1$ is excluded from the parameter range, we actually have
\begin{align}
-1\le ab< 1\,.
\label{ab2}
\end{align}

\subsection{Geometrical and physical properties}

\subsubsection{Rod structure}

We now consider the rod structure of the rotating black bottle solution. As in the static case, useful information about the geometrical and physical properties of the space-time can be deduced from it. We begin with the rod structure of the rotating black bottle with an acceleration horizon:
\begin{itemize}
\item Rod 1: a semi-infinite space-like rod located at $(x\,{=}\,{-1},y_1\,{\leq}\,y\,{<}\,{-1})$, with direction
    \begin{align}
    k_1=\frac{1}{\kappa_{\text{E}1}}(\sqrt{a},1)\,,\qquad \kappa_{\text{E}1}=0\,;
\end{align}

\item Rod 2: a finite time-like rod located at $(-1\,{\le}\,x\,{\le}\,{+1},y\,{=}\,y_1)$, with direction
\begin{align}
\label{rod2}
k_2=\frac{1}{\kappa_2}(1,-\sqrt{a}y_1^2)\,,\qquad \kappa_2=-\frac{1}{2}\frac{\dif Q}{\dif y}\bigg|_{y=y_1}\,;
\end{align}

\item Rod 3: a finite space-like rod located at $(x\,{=}\,{+1},y_1\,{\le}\,y\,{\le}\,y_2)$, with direction
\begin{align}
k_3=\frac{1}{\kappa_{\text{E}3}}(\sqrt{a},1)\,,\qquad \kappa_{\text{E}3}=2(1+ab)\,;
\end{align}

\item Rod 4: a semi-infinite time-like rod located at $(y_2\,{<}\,x\,{\le}\,{+1},y\,{=}\,y_2)$, with direction
\begin{align}
\label{rod4}
k_4=\frac{1}{\kappa_4}(1,-\sqrt{a}y_2^2)\,,\qquad \kappa_4=\frac{1}{2}\frac{\dif Q}{\dif y}\bigg|_{y=y_2}\,.
\end{align}
\end{itemize}
The rod structure of the rotating black bottle without an acceleration horizon can be obtained from this by simply removing the fourth rod. Without this rod, Rod 3 will extend to conformal infinity $x=y$; it now has coordinates $(x\,{=}\,{+1},y_1\,{\leq}\,y\,{<}\,{+1})$.

Note that in both cases, the locations of the rods and the turning points between them are formally identical to those of the static case. The introduction of rotation, however, changes the directions of all the rods. In particular, their directions are no longer purely along $\frac{\partial}{\partial t}$ or $\frac{\partial}{\partial\phi}$. This necessitates the introduction of new temporal and/or azimuthal coordinates, which we will do in the following subsection.

\subsubsection{Temporal and azimuthal coordinates}

If we define a new temporal coordinate $\tau$ by
\begin{align}
\tau=t-\sqrt{a}\phi\,,
\label{linear_transformation}
\end{align}
the metric (\ref{metric_rotating_droplet}) becomes
\begin{align}
\dif s^2&=\frac{\ell^2(1-b)}{(x-y)^2}\bigg\{\frac{Q}{1+a x^2y^2}\big[\dif \tau+\sqrt{a}(1-x^2)\dif\phi\big]^2-\frac{1+a x^2y^2}{Q}\,\dif y^2\nonumber\\
&\hspace{0.788in}+\frac{1+a x^2y^2}{P}\,\dif x^2+\frac{P}{1+a x^2y^2}\big[(1+ay^2)\dif \phi+\sqrt{a}y^2\dif\tau\big]^2\bigg\}\,.
\label{metric_rotating_droplet_new_coordinates}
\end{align}
One can then recalculate the rod structure in these coordinates. The transformation (\ref{linear_transformation}) actually does not change the locations of the rods and turning points found in the preceding subsection. The rod directions are also invariant, but they should now be expressed in the new basis $\big\{\frac{\partial}{\partial \tau},\frac{\partial}{\partial\phi}\big\}$ \cite{Chen:2010zu}. They are given by
\begin{subequations}
\label{rod_structure_new_coordinates}
\begin{align}
k_1&=\frac{1}{\varkappa_{\text{E}1}}(0,1)\,,& &\hskip-2cm\varkappa_{\text{E}1}=0\,;\\
\label{rod_structure_new_coordinates_2}
k_2&=\frac{1}{\varkappa_2}\left(1,-\frac{\sqrt{a}y_1^2}{1+ay_1^2}\right),& &\hskip-2cm\varkappa_2=-\frac{1}{2(1+ay_1^2)}\frac{\dif Q}{\dif y}\bigg|_{y=y_1}\,;\\
k_3&=\frac{1}{\varkappa_{\text{E}3}}(0,1)\,,& &\hskip-2cm\varkappa_{\text{E}3}=2(1+ab)\,;\\
\label{rod_structure_new_coordinates_4}
k_4&=\frac{1}{\varkappa_4}\left(1,-\frac{\sqrt{a}y_2^2}{1+ay_2^2}\right),& &\hskip-2cm\varkappa_4=\frac{1}{2(1+ay_2^2)}\frac{\dif Q}{\dif y}\bigg|_{y=y_2}\,.
\end{align}
\end{subequations}
We remind the reader that $k_4$ is irrelevant in the case of the black bottle without an acceleration horizon.

In these coordinates, it is clear that Rod 3 is an axis of the space-time, with $\phi$ as the azimuthal coordinate. To avoid a conical singularity along this axis, the coordinate identification
\begin{align}
(\tau,\phi)\rightarrow \Big(\tau,\phi+\frac{\pi}{1+ab}\Big)\,,
\label{identification}
\end{align}
has to be made.

As in the static case, the vanishing of $\varkappa_{\text{E}1}$ indicates that Rod 1 represents a new spatial infinity of the space-time. The fact that $k_1$ is parallel to $k_3$ implies that it is the $\phi$-circle which vanishes at the new spatial infinity.

\subsubsection{Horizon geometries}

Since Rod 2 and Rod 4 (if it exists) are time-like, they represent horizons in the space-time. They are the rotating generalisations of the black-hole and acceleration horizons respectively. We now turn to a study of their geometries. 

We begin with the black-hole horizon represented by Rod 2. It is convenient to reparameterise the solution in terms of $a$ and $y_1$, which amounts to writing $b$ in terms of $a$ and $y_1$ as
\begin{align}
b=\frac{y_1(-1+y_1+y_1^2+ay_1^3)}{1+a(y_1+y_1^2-y_1^3)}\,.
\label{b_in_y1}
\end{align}
For a constant time slice, the horizon has the induced metric
\begin{align}
\label{induced}
\dif s_{\text{BH}}^2&=\frac{\ell^2(1-b)}{(x-y_1)^2}\bigg[\frac{1+a x^2y_1^2}{P}\,\dif x^2+\frac{P}{1+a x^2y_1^2}(1+ay_1^2)^2\dif \phi^2\bigg]\,.
\end{align}
With the form of $P$ in (\ref{structure_functions_rotating}), one can check that the geometry around the north pole, $x=+1$, resembles that around the north pole of a perfect sphere. On the other hand, at the south pole, $x=-1$, the horizon extends to the new spatial infinity in the form of a cusp. 

The scalar curvature of the metric (\ref{induced}) can be calculated to be
\begin{align}
R=-\frac{2}{\ell^2}+\frac{2(x-y_1)^3(1+ay_1^2)^2(1-3axy_1-3ax^2y_1^2+a^2x^3y_1^3)}{\ell^2(1-y_1)(1+y_1)^2(1+ay_1)(1+ax^2y_1^2)^3}\,.
\end{align}
At the north and south poles, and at the equator, it is
\begin{subequations}
\begin{align}
\label{R_3}
R(x\,{=}\,{+1})&=-\frac{8y_1(1+a)(1+ay_1^3)}{\ell^2(1+y_1)^2(1+ay_1)(1+ay_1^2)}\,,\\
\label{R_1}
R(x\,{=}\,{-1})&=-\frac{4(1+2ay_1-2ay_1^3-a^2y_1^4)}{\ell^2(1-y_1)(1+ay_1)(1+ay_1^2)}<0\,,\\
\label{R_2}
R(x\,{=}\,0)&=-\frac{2[(1+ay_1)(1+y_1-y_1^2)-ay_1^4(1-2y_1-ay_1^3)]}{\ell^2(1-y_1)(1+y_1)^2(1+ay_1)}>0\,.
\end{align}
\end{subequations}
It can be shown that the scalar curvature at the south pole (\ref{R_1}) is always negative, while that at the equator (\ref{R_2}) is always positive. Thus, the geometry is hyperbolic at the south pole, consistent with the formation of a cusp there. The geometry is spherical around the equator. However, the scalar curvature at the north pole (\ref{R_3}) can either be positive or negative. It can be shown that it is positive for sufficiently small $a$, and negative for sufficiently large $a$. 

It follows that, for sufficiently small $a$, the horizon interpolates between a spherical geometry at the north pole and a hyperbolic one at the south pole. This is similar to the situation in the static case, and the horizon has the shape of a bottle. Indeed, for such cases, embeddings of (\ref{induced}) in a three-dimensional Euclidean space are similar to that in Fig.~\ref{fig_bottle}.

On the other hand, for sufficiently large $a$, the north pole of the horizon has a hyperbolic geometry. For such cases, it turns out that an embedding of (\ref{induced}) in a three-dimensional Euclidean space is not possible, at least near the north pole. (The south pole up to the equatorial region can still be embedded, and would look like the corresponding regions in Fig.~\ref{fig_bottle}.) Nonetheless, the horizon will in general still have the topology of a sphere with one puncture, i.e., a bottle. For this reason, we will continue to refer to these solutions as black bottles. It is only in the black spindle limit (c.f.\ Sec.~\ref{sec_spindle}) that the north pole will extend to infinity, and the horizon topology will become that of a sphere with two punctures.

Recall from the rod structure (\ref{rod_structure_new_coordinates}) that the directions of Rods 1 and 3 are parallel. This means that the generator of the neck of the black bottle, i.e., the direction that circles around the neck of the bottle, is also the generator of the axis of the space-time. So the bottle is rotating about its neck. Moreover, it rotates with angular velocity $\Omega_{\rm BH}$, given by the second component in the bracket of $k_2$ in (\ref{rod_structure_new_coordinates_2}).

The area of the horizon is finite as in the static case, and can be straightforwardly calculated to be
\begin{align}
A_{\text{BH}}=-\frac{2\pi\ell^2(1+y_1)(1+ay_1)}{1+ay_1^2}\,.
\end{align}
We see that, for fixed $y_1$, increasing the rotational parameter $a$ decreases the area of the horizon. The minimal area occurs when $a$ reaches the curve (a) in Fig.~\ref{fig_parameter_space}, corresponding to an extremal black bottle (c.f.\ Sec.~\ref{sec_ext_bottle}).

Turning to the acceleration horizon represented by Rod 4, we can similarly calculate its induced metric:
\begin{align}
\label{induced2}
\dif s_{\text{AH}}^2=\frac{\ell^2(1-b)}{(x-y_2)^2}\bigg[\frac{1+a x^2y_2^2}{P}\,\dif x^2+\frac{P}{1+a x^2y_2^2}(1+ay_2^2)^2\dif \phi^2\bigg]\,.
\end{align}
It can be checked that the geometry is regular around the north pole, $x=+1$. At the other end of the range, $x=y_2$, the horizon extends to conformal infinity of the space-time.

The scalar curvature of the metric (\ref{induced2}) can be calculated, and it can be seen that it is always negative at $x=y_2$. However, the scalar curvature at $x=+1$ can take either sign in general, as in the static case. If it is positive, a partial embedding of the horizon is possible, and this embedding is similar to the droplet-shaped horizon in Fig.~\ref{fig_droplet}.

The second component in the bracket of $k_{4}$ in (\ref{rod_structure_new_coordinates_4}) represents the angular velocity $\Omega_{\rm AH}$ of the acceleration horizon. From the fact that $y_1^2>y_2^2$, we have for $a\neq0$,
\begin{align}
|\Omega_{\text{BH}}|>|\Omega_{\text{AH}}|\,,
\end{align}
i.e., the bottle horizon rotates faster than the acceleration horizon. In particular, they can never be in dynamical equilibrium. We remark that by appropriately choosing the time coordinate, one can find a coordinate system in which the acceleration horizon is static at conformal infinity. However, the bottle horizon will still have a non-zero angular velocity.

\subsubsection{Absence of CTCs}

We now show that the rotating black bottle space-time does not contain any closed time-like curves (CTCs). The absence of CTCs in a space-time with axial symmetry requires that the generator of this symmetry never vanishes except on the axis itself. In other words, $\frac{\partial}{\partial\phi}$ in the metric (\ref{metric_rotating_droplet_new_coordinates}) should have a non-negative norm $g_{\phi\phi}$ in the domain of interest. Direct computation yields
\begin{align}
g_{\phi\phi}=\frac{\ell^2(1-b)(1+ab)(1-x)(1+x)^2H}{(x-y)^2(1+a x^2y^2)}\,,
\end{align}
where $H$ is defined as
\begin{align}
H\equiv(1-axy)(1-ay^3)+ay(1+y)(1+xy)\,.
\end{align}

We now prove that $H$ is positive in the region
\begin{align}
-1<x<+1\,,\qquad  y_1<y<x\,;
\end{align}
this is sufficient to rule out CTCs in the two possible domains (\ref{domain1}) and (\ref{domain2}), corresponding to black bottles without and with acceleration horizons respectively. Recall that the parameters $a$ and $b$ obey the constraints (\ref{ab}) and (\ref{ab2}). Here, for simplicity, we shall assume $0<a<1$; it can be verified directly that the static case $a=0$ has no CTCs. 

We begin by noting that $H$ is linear in terms of $x$. Firstly, we consider the triangular range of coordinates satisfying
\begin{align}
-1\le y<x<+1\,.
\end{align}
In this range, we clearly have $1+ay>0$ and $1+ay^4>0$. $H$ is then positive in this range since
\begin{align}
H(x\,{=}\,y)=(1+ay)(1+ay^4)>0\,,\qquad H(x\,{=}\,{+1})=(1+ay^2)^2>0\,.
\end{align}

Next, we consider the rectangular range of coordinates satisfying
\begin{align}
y_1<y<-1<x<+1\,.
\end{align}
Since we have $H(x\,{=}\,{+1})>0$, we need to prove
\begin{align}
K\equiv H(x\,{=}\,{-1})=1+2ay-2ay^3-a^2y^4>0\,,
\end{align}
for all $y_1<y<-1$. We observe that $K$ is a quartic polynomial in terms of $y$ with negative leading coefficient, and that
\begin{align}
K'(y\,{=}\,{-1})=4a(a-1)<0\,,\qquad K'(y\,{=}\,0)=2a>0\,.
\end{align}
This shows that there is exactly one maximum for $K$ lying in the range $-\infty<y<-1$. So for $y_1<y<-1$, the {\it minimum\/} of $K$ lies at either endpoint of the range. In either case, $K$ is positive, as can be seen as follows:
\begin{align}
K(y\,{=}\,{-1})&=1-a^2>0\,,\nonumber\\
K(y\,{=}\,y_1)&=K(y\,{=}\,y_1)-aQ(y\,{=}\,y_1)\nonumber\\
&=(1-ab)\big[1+ay_1^2+ay_1(1-y_1^2)\big]>0\,.
\end{align}
This concludes the proof that there are no CTCs in the domains of interest.

\subsection{Special cases}
\label{sec_limits}

In this section, we shall study several important special cases of the rotating black bottle solution (\ref{metric_rotating_droplet}) or (\ref{metric_rotating_droplet_new_coordinates}). In doing so, we shall come to a physical understanding of the various boundaries and parts of the parameter space in Fig.~\ref{fig_parameter_space}.

\subsubsection{Extremal black bottle}
\label{sec_ext_bottle}

As mentioned above, the rotating black bottle has two horizons: an outer one at $y=y_1$ and an inner one at $y=y_0$. An extremal black bottle occurs when the two horizons coincide:
\begin{align}
y_0=y_1\,.
\end{align}
This condition is most elegantly solved if we adopt $y_1$ as a fundamental parameter; $a$ and $b$ can then be expressed in terms of it as
\begin{align}
\label{curve(a)}
a=\frac{1-3y_1}{y_1^2(3-y_1)}\,,\qquad b=-\frac{y_1^2(3+y_1)}{1+3y_1}\,,\qquad y_1<-1\,.
\end{align}
This gives the curve (a) in Fig.~\ref{fig_parameter_space}. Being the upper boundary of the parameter space, this curve yields the maximum value of $a$ possible for a given value of $b$.

In terms of $y_1$, the other two roots of $Q$ can be written as
\begin{align}
y_{2,3}=\frac{y_1\big(1-y_1^2\pm 2\sqrt{-2+20y_1^2-2y_1^4}\big)}{9y_1^2-1}\,.
\end{align}
It can be checked that $y_{2,3}$ are complex in the range 
\begin{align}
y_1<-\sqrt{3}-\sqrt{2}\,.
\end{align}
In this case, the solution describes an extremal black bottle without an acceleration horizon. On the other hand, $y_{2,3}$ are real in the range
\begin{align}
-\sqrt{3}-\sqrt{2}\leq y_1<-1\,,
\end{align}
and the solution describes an extremal black bottle with an acceleration horizon. The critical value $y_1=-\sqrt{3}-\sqrt{2}$ corresponds to the case in which $y_2=y_3$. In this case, we have
\begin{align}
a=-b=\sqrt{9}-\sqrt{8}\,,
\end{align}
and the four roots of $Q$ are
\begin{align}
y_0=y_1=-\sqrt{3}-\sqrt{2}\,,\qquad y_2=y_3=\sqrt{3}-\sqrt{2}\,.
\end{align}
This special case, which is indicated by a bullet point in Fig.~\ref{fig_parameter_space}, describes an extremal black bottle with an extremal acceleration horizon.

\subsubsection{Black spindle}
\label{sec_spindle}

Recall from (\ref{structure_functions_rotating}) that the function $P$ has four real roots: a double root at $x=-1$, a single root at $x=+1$, and another single root at $x=-\frac{1}{ab}$. By the condition (\ref{ab2}), the last root lies outside the physical range $-1<x<+1$. However, in the special case 
\begin{align}
\label{ab=-1}
ab=-1\,,
\end{align}
it will coincide with the root at $x=+1$ to form a double root there. We then have the interesting situation in which the physical range of $x$ is bounded by double roots at {\it both\/} endpoints.

It can be checked that for fixed values of $b<0$, the curve (a) in Fig.~\ref{fig_parameter_space} always lies below that of (\ref{ab=-1}). It is only in the limit $b\rightarrow-\infty$ that they coincide, at the extreme left edge of the parameter space in Fig.~\ref{fig_parameter_space}. Thus the special case (\ref{ab=-1}) can only be achieved in the limit $b\rightarrow-\infty$. At the same time, we need to take $a\rightarrow0$ so that (\ref{ab=-1}) is satisfied.

To take this limit in an appropriate way, we first set
\begin{align}
b=-\frac{1}{\epsilon^2}\,,\qquad a=\epsilon^2(1-c\epsilon)\,,\qquad y=-\frac{1}{\epsilon r}\,,\qquad t\rightarrow \epsilon t\,,
\label{limit_spindle}
\end{align}
and then take the limit $\epsilon\rightarrow 0^+$. The metric (\ref{metric_rotating_droplet}) then becomes
\begin{align}
\dif s^2&={\ell}^{2} \bigg[ -{\frac {F ( \dif{{t}}-{x}^{2}\dif{{\phi}} )^2}{r^2+x^2}}+{\frac { (r^2+x^2) {\dif{r}}^2}{F}}+{\frac {(r^2+x^2){\dif{x}}^2}{G}}+{\frac {G (\dif{t}+r^2\dif{\phi})^{2}}{r^2+x^2}} \bigg]\,,\nonumber\\
	G&=(1-x^2)^2,\qquad F=(1+r^2)^2-cr\,.
\end{align}
This solution has a curvature singularity at $(r\,{=}\,0,x\,{=}\,0)$. For it to describe a black hole, we require that $F$ has a positive root representing the location of an event horizon. This implies that
\begin{align}
c\ge \frac{16\sqrt{3}}{9}\,.
\label{range_spindle}
\end{align}
It can be checked that for such values of $c$, and only such values, the limit (\ref{limit_spindle}) falls within the parameter ranges that we have identified. The case of equality in (\ref{range_spindle}) describes an extremal black hole. This solution was first identified by Klemm \cite{Klemm:2014rda}.

The defining property of this solution is that $P$ has a pair of double roots, at $x=\pm1$. As we have seen in the black bottle solution, a double root of $P$ corresponds to a new spatial infinity of the space-time. The horizon extends to this new spatial infinity in the form of a cusp. It follows that in this case, the horizon will have two separate cusps, extending in opposite directions to the new spatial infinities at $x=\pm1$, as in Fig.~\ref{fig_spindle}. The black bottle becomes a black spindle.

Unlike the black bottle however, the present solution does not contain any space-like rod with non-vanishing surface gravity. This means that the space-time does not contain any axis, and we no longer have to make an identification on the azimuthal coordinate $\phi$ to ensure that the space-time is regular. A consequence of this is that $\phi$ can take any period, or can even be uncompactified, i.e., $-\infty< \phi<\infty$. We remark that if $\phi$ has a finite period, the area of the horizon is finite.

\subsubsection{Black bottle with an extremal acceleration horizon}

Now we focus on the case in which the black bottle has an acceleration horizon. Besides $y_{0,1,2}$, the function $Q$ has a fourth real root $y_3$. This root lies beyond the acceleration horizon at $y=y_2$, and so is outside the physical range for $y$. However, in the special case 
\begin{align}
\label{y2=y3}
y_2=y_3\,,
\end{align}
it will coincide with the root at $y=y_2$ to form a double root there. This corresponds to the acceleration horizon becoming extremal.

The condition (\ref{y2=y3}) can be solved in terms of $y_2$ as follows:
\begin{align}
a=\frac{1-3y_2}{y_2^2(3-y_2)}\,,\qquad b=-\frac{y_2^2(3+y_2)}{1+3y_2}\,,\qquad \sqrt{3}-\sqrt{2}\leq y_2\leq \frac{1}{3}\,.
\label{range_y2_droplet_ex_accl}
\end{align}
Note that this solution has the same form as the one in (\ref{curve(a)}), although the parameter range is, of course, different. The range of $y_2$ in (\ref{range_y2_droplet_ex_accl}) is restricted as such by the requirement that the roots $y_{0,1}$ are real, so that the inner and outer horizons of the black bottle exist. In terms of $y_2$, these two roots can be written as
\begin{align}
y_{0,1}=\frac{y_2\big(1-y_2^2\pm 2\sqrt{-2+20y_2^2-2y_2^4}\big)}{9y_2^2-1}\,.
\end{align}
The lower bound of $y_2$ in (\ref{range_y2_droplet_ex_accl}) corresponds to the case $y_0=y_1$. In this case, the solution (\ref{metric_rotating_droplet_new_coordinates}) describes an extremal black bottle with an extremal acceleration horizon, a configuration which was discussed in Sec.~\ref{sec_ext_bottle}. On the other hand, the upper bound of $y_2$ in (\ref{range_y2_droplet_ex_accl}) corresponds to the case $a=0$, i.e., the static limit of the black bottle with an extremal acceleration horizon.

In the parameter space of Fig.~\ref{fig_parameter_space}, (\ref{range_y2_droplet_ex_accl}) gives the dashed curve. This curve divides the parameter space into two regions: on its right are the black bottles with acceleration horizons, characterised by $Q$ having four real roots; on its left are the black bottles without acceleration horizons, characterised by $Q$ having only two real roots.

\subsubsection{Black bottle in thermal equilibrium with its acceleration horizon}

In the case when the black bottle has an acceleration horizon, each horizon has its own temperature, and it is natural to ask if thermal equilibrium is possible in such a configuration. Since the temperature of a horizon is proportional to its surface gravity, the condition of thermal equilibrium in this case translates to the equality $\varkappa_2=\varkappa_4$. From (\ref{rod_structure_new_coordinates_2}) and (\ref{rod_structure_new_coordinates_4}), we thus have the condition
\begin{align}
\frac{1}{1+ay_1^2}\frac{\dif Q}{\dif y}\bigg|_{y=y_1}+\frac{1}{1+ay_2^2}\frac{\dif Q}{\dif y}\bigg|_{y=y_2}=0\,.
\label{thermal_equi}
\end{align}
This equation involves the roots $y_1$ and $y_2$ explicitly, suggesting that it might be easier to solve using them as parameters. The parameterisation of $a$ and $b$ in terms of $y_{1,2}$ is then given by the equations
\begin{subequations}
\begin{align}
Q(y_1)=0\,,\label{Qy1}\\
\quad Q(y_2)=0\,.
\label{Qy2}
\end{align}
\end{subequations}

The equation (\ref{Qy1}) is solved by (\ref{b_in_y1}). Substituting this into (\ref{Qy2}) and (\ref{thermal_equi}), we obtain two polynomial equations in terms of $a$ and $y_{1,2}$. These two equations are quadratic in terms of $a$. A linear combination of them that is linear in $a$ can then be found by eliminating the $a^2$ term. Solving this equation for $a$ in the form $a=a(y_1,y_2)$, and then substituting back into (\ref{Qy2}) and (\ref{thermal_equi}), we get the relation
\begin{align}
2(1-y_1y_2)^2-(1+y_1+y_2+y_1y_2)^2=0\,.
\end{align}
A solution for $y_2$ is given by
\begin{align}
y_2=-\frac{y_1+1+\sqrt{2}}{1+(1-\sqrt{2})y_1}\,.
\label{y2_in_y1}
\end{align}
The other solution for $y_2$ is discarded since it leads to an incorrect signature for the domain of interest.

The solution is thus completely determined by the parameters $\ell$ and $y_1$. Substituting (\ref{y2_in_y1}) into $a(y_1,y_2)$ and (\ref{b_in_y1}), we obtain the parameters $a$ and $b$ in terms of $y_1$ as follows:
\begin{align}
 a=\sqrt{9}-\sqrt{8}\,,\qquad 
b=\frac{y_1(y_1+1+\sqrt{2})(y_1^2+2y_1+\sqrt{2}y_1-1-\sqrt{2})}{(-y_1+1+\sqrt{2})(y_1^2+\sqrt{2}y_1+1+\sqrt{2})}\,,
\end{align}
for $-\sqrt{3}-\sqrt{2}\leq y_1<-1$. The black bottle in thermal equilibrium with its acceleration horizon is thus described in the $a$-$b$ parameter space by the straight line
\begin{align}
\label{curve(c)}
a=\sqrt{9}-\sqrt{8}\,,\qquad \sqrt{8}-\sqrt{9}\leq b<1\,.
\end{align}
The lower bound of $b$ in (\ref{curve(c)}) corresponds to the special case in which thermal equilibrium is achieved at zero temperature, when the two horizons become extremal.

The curve (\ref{curve(c)}) is the dotted one in Fig.~\ref{fig_parameter_space}. It divides the region of the parameter space describing black bottles with acceleration horizons into two subregions: above it are configurations in which the bottle is colder than the acceleration horizon (the so-called ``cold bottles''); below it are configurations in which the bottle is hotter than the acceleration horizon (the so-called ``hot bottles'').

\section{Summary and discussion}

In this paper, we have presented a new solution, (\ref{metric_rotating_droplet}) or (\ref{metric_rotating_droplet_new_coordinates}), describing a rotating black bottle in an asymptotically AdS space-time. Besides the AdS length scale $\ell$, the solution has two other parameters $a$ and $b$, which fill out the region of the parameter space as shown in Fig.~\ref{fig_parameter_space}. $a$ can be interpreted as a rotation parameter, with the static black bottle recovered when $a=0$. On the other hand, $b$ can be interpreted as an acceleration parameter. When it is sufficiently small, the space-time contains only the black bottle horizon; otherwise, an extra acceleration horizon appears in the space-time. 

The latter behaviour is also known to occur in the case of accelerating spherical black holes in AdS space-time \cite{Podolsky:2002nk,Dias:2002mi,Krtous:2005ej}. When the acceleration of the black hole is sufficiently small, it actually remains static with respect to AdS infinity. In this case, a conical singularity attached to the black hole provides the necessary tension to counterbalance the cosmological compression of AdS space. In the black bottle case however, it is the neck of the bottle which plays the role of the conical singularity in connecting the black bottle to AdS infinity and keeping it static.

We remark that the acceleration is turned off in the limit $b\rightarrow-\infty$. This is the limit in which we can recover the black spindle, so the latter can be interpreted as a black bottle with vanishing acceleration. This is consistent with the fact that the black spindle can be obtained from the Carter--Pleba\'nski solution, which is the zero-acceleration limit of the Pleba\'nski--Demia\'nski solution. It is also consistent with the fact that the black spindle can be obtained as an ultra-spinning limit of the non-accelerating Kerr--AdS black hole. 

Like the black spindle, one of the most interesting properties of the black bottle is that it has a non-compact event horizon with a finite area. It follows that its entropy is also finite. For the black spindle, this results in some rather unusual thermodynamic properties. It was shown in \cite{Hennigar:2014cfa,Hennigar:2015cja} that the black spindle provides the first counterexample to the so-called reverse isoperimetric inequality. This is a conjecture that amongst black holes of a given thermodynamic volume, the Schwarzschild--AdS black hole has the maximal entropy. It turns out that the black spindle exceeds the maximal entropy implied by the reverse isoperimetric inequality, and for this reason, it was referred to as ``super-entropic'' in \cite{Hennigar:2014cfa,Hennigar:2015cja}. It would be interesting to check if the black bottle also violates the reverse isoperimetric inequality.

The static solution describing a black bottle with an acceleration horizon has previously been studied within the context of the AdS/CFT correspondence \cite{Hubeny:2009ru}. Since the acceleration horizon (known as a black droplet in \cite{Hubeny:2009ru}) extends to conformal infinity $x=y$, there is a black hole on this boundary where the CFT is defined. It is then possible to use the AdS/CFT correspondence to study the dynamics of a strongly coupled field theory in the background of this boundary black hole. In the AdS bulk, the interaction between the field theory plasma and the boundary black hole can be understood in terms of the interaction between the two bulk horizons, namely the black bottle and the black droplet. 

Since the focus in \cite{Hubeny:2009ru} was on the description of equilibrium states, the black bottle and black droplet horizons were supposed to have the same temperature. However, in the static solution considered therein, the two horizons always had different temperatures. In \cite{Caldarelli:2011wa}, the authors added charge to this solution, and found that thermal equilibrium was possible for a certain set of parameters. In this paper, we have seen that adding rotation is another way to achieve thermal equilibrium. But we have also noted that since the two horizons always have different angular velocities, dynamical equilibrium is not possible in the rotating solution.

A different static black-hole solution, known as a black funnel, was also considered in \cite{Hubeny:2009ru} as a possible AdS bulk. In this solution, there is a {\it single\/} horizon stretching from conformal infinity all the way to the new spatial infinity. This solution is, in fact, again given by (\ref{metric_static_droplet}), with the parameter range (\ref{range_static_droplet}). However, the coordinate range is now different. In the context of Fig.~\ref{fig_domain_static_droplet}, the appropriate domain is the triangle which is bounded by the lines $x=-1$, $y=y_2$ and $x=y$. In particular, the line $y=y_2$ is the black funnel horizon. The solution describing a rotating black funnel can similarly be read off from (\ref{metric_rotating_droplet}), with the appropriate coordinate and parameter ranges.

Although the black bottles considered in this paper are uncharged, it is possible to add an electric charge $e$ and a magnetic charge $g$ to them. The charged version of (\ref{metric_rotating_droplet}) is given by
\begin{subequations}
\label{metric_charged_droplet}
\begin{align}
	\dif s^2&=\frac{\ell^2(1-b)}{(x-y)^2}\bigg[\frac{Q}{1+a x^2y^2}(\dif t-\sqrt{a}x^2\dif \phi)^2-\frac{1+a x^2y^2}{Q}\,\dif y^2
\nonumber\\
	&\hspace{0.74in}+\frac{1+a x^2y^2}{P}\,\dif x^2+\frac{P}{1+a x^2y^2}(\dif \phi+\sqrt{a}y^2\dif t)^2\bigg]\,,\nonumber\\
	P&=1-q^2+(1+ab)x-(1-ab-2q^2)x^2-(1+ab)x^3-(ab+q^2)x^4,\nonumber\\
	Q&=b-q^2+(1+ab)y-(1-ab-2q^2)y^2-(1+ab)y^3-(a+q^2)y^4,
\end{align}
and the corresponding gauge potential is 
\begin{align}
{\cal A}=\frac{\sqrt{\ell^2(1-b)(1+a)}}{1+ax^2y^2}\,\big[ey(\dif t-\sqrt{a}x^2\dif\phi)-gx(\dif \phi+\sqrt{a}y^2\dif t)\big]\,.
\end{align}
\end{subequations}
Here, we have defined $q\equiv\sqrt{e^2+g^2}$. Note that $P$ and $Q$ can also be written as
\begin{align}
P(x)=(1-x)(1+x)^2\big[1-q^2+(ab+q^2)x\big]\,,\qquad Q(y)=P(y)-(1-b)(1+ay^4)\,.
\end{align}
This solution describes a charged rotating black bottle. If we set $a=0$, we recover the static charged black bottle solution first studied in \cite{Caldarelli:2011wa}. In Appendix~\ref{sec_appendix}, we show how (\ref{metric_charged_droplet}) can be derived from the general Pleba\'nski--Demia\'nski solution. It may be worthwhile to study this solution in more detail.

Finally, we mention that the solution (\ref{metric_rotating_droplet}) admits a natural generalisation in which $P$ no longer has a double root at $x=-1$. Thus, the south pole of the horizon would no longer be the end of a cusp that is infinitely far away. At the same time, the sign of the cosmological constant is no longer restricted to be negative. Such a four-parameter solution would describe a {\it spherical\/} black hole in either a de Sitter or an anti-de Sitter space-time, that possesses both a rotation and an acceleration, but without a NUT charge. It is in fact a generalisation of the rotating C-metric solution of \cite{Hong:2004dm} to include a cosmological constant. We will present the new form of this solution and analyse its properties in a forthcoming publication \cite{C_rotating}.

\section*{Acknowledgement}

This work was supported by the Academic Research Fund (WBS No.: R-144-000-333-112) from the National University of Singapore.

\appendix

\section{Derivation of the black bottle solution}
\label{sec_appendix}

In this appendix, we show how the various black bottle solutions that we have studied can be derived from the general Pleba\'nski--Demia\'nski solution.

The Pleba\'nski--Demia\'nski solution \cite{Plebanski:1976gy} has the metric:
\begin{subequations}
\label{PD}
\begin{align}
\dif s^2&=\frac{1}{(p-q)^2}\bigg[\frac{Q}{1+p^2q^2}(\dif\tau-p^2\dif\phi)^2-\frac{1+p^2q^2}{Q}\,\dif q^2\nonumber\\
&\hspace{0.69in}+\frac{1+p^2q^2}{P}\,\dif p^2+\frac{P}{1+p^2q^2}(\dif\phi+q^2\dif\tau)^2\bigg]\,,\nonumber\\
P&=\gamma_1+2np-\epsilon p^2+2mp^3-(\gamma_2+e^2+g^2) p^4,\nonumber\\
Q&=\gamma_2+2nq-\epsilon q^2+2mq^3-(\gamma_1+e^2+g^2) q^4,
\end{align}
and the gauge potential:
\begin{align}
{\cal A}=\frac{1}{1+p^2q^2}\left[eq(\dif t-p^2\dif \phi)-gp(q^2\dif t+\dif\phi)\right].
\end{align}
\end{subequations}
It can be checked that (\ref{PD}) is a solution to the Einstein--Maxwell equations, with cosmological constant $\Lambda=3(\gamma_2-\gamma_1)$. The parameters $e$ and $g$ are related to the electric and magnetic charge of the solution respectively. Note that for fixed $\Lambda$, $e$ and $g$, the solution is completely determined by the function $P$. 

The derivation starts from the observation that to have a bottle geometry, $P$ should possess a double root at the neck of the bottle, which we assume to be at $p=p_1$. The Killing vector that circles around the neck is then given by
\begin{align}
k_{1}\propto p_{1}^2\frac{\partial}{\partial t}+\frac{\partial}{\partial\phi}\,.
\end{align}
On the other hand, the generator of the axis, which we assume to be at $p=p_2$, is given by
\begin{align}
k_{2}\propto p_{2}^2\frac{\partial}{\partial t}+\frac{\partial}{\partial\phi}\,.
\end{align}
We demand that these two vectors are parallel, so that the space-time is rotating about the neck. This implies that $p_1^2=p_2^2$, and the only non-trivial solution to this is $p_1=-p_2$. Hence, to have a black bottle solution, we require that $P$ has a pair of opposite roots ($-p_2$ and $p_2$), with one of them ($-p_2$) being degenerate.

The above conditions on $P$ can be summarised as $P(p_2)=P(-p_2)=P'(-p_2)=0$, which can then be solved to obtain
\begin{align}
\gamma_1&=\frac{p_2^4(1-e^2\ell^2-g^2\ell^2)-2m\ell^2p_2^3}{\ell^2(1+p_2^4)}\,,\qquad n=-mp_2^2\,,\nonumber\\ \epsilon&=\frac{2p_2^2(1-e^2\ell^2-g^2\ell^2)-2m\ell^2p_2(1-p_2^4)}{\ell^2(1+p_2^4)}\,,
\end{align}
where $\ell$ is related to the cosmological constant $\Lambda$ by (\ref{Lambda}). Substituting these expressions into $P$, we obtain
\begin{align}
P=\frac{(p-p_2)(p+p_2)^2[(p-p_2)(1-e^2\ell^2-g^2\ell^2)+2m\ell^2(1+p_2^3p)]}{\ell^2(1+p_2^4)}\,.
\end{align}
This parameterisation of $P$ in terms of $p_2$ and $m$, in addition to $\ell$, $e$ and $g$, uniquely determines the black bottle solution. There are five independent parameters in all.

The remaining task is to simplify the metric. Firstly, we would like to rescale the coordinates so that the roots of $P$ now lie at $\pm1$. Secondly, we want $P$ to become a cubic polynomial when $p_2\rightarrow 0$, which corresponds to taking the static limit. These considerations motivate us to perform the rescalings:
\begin{align}
p&\rightarrow p_2x\,,\qquad q\rightarrow p_2y\,,\qquad t\rightarrow p_2t\,,\qquad \phi\rightarrow p_2\phi\,,\cr
m&\rightarrow \frac{m}{\ell^2p_2^3}\,,\qquad (e,g)\rightarrow \frac{(e,g)}{\ell p_2^2}\,,
\end{align}
which results in a $P$ of the form:
\begin{align}
P=\frac{(x-1)(x+1)^2[2m(1+xp_2^4)+(1-x)(e^2+g^2-p_2^4)]}{\ell^2(1+p_2^4)}\,.
\end{align}
In the static limit $p_2=0$, and with $e=g=0$, the expression of $P$ in (\ref{structure_functions_static}) is indeed obtained (up to an overall factor). We note that $p_2$ appears only in the form $p_2^4$. So we redefine $p_2$ and $m$ in terms of new parameters $a$ and $b$ as follows:
\begin{align}
p_2=\sqrt[4]{a}\,,\qquad m=\frac{1+ab}{2(b-1)}\,.
\end{align}
The final simplification comes from rescaling $e$ and $g$ by
\begin{align}
(e,g)\rightarrow\sqrt{\frac{1+a}{1-b}}\,(e,g)\,,
\end{align}
which results in a very simple form for $P$:
\begin{align}
P=\frac{(x-1)(x+1)^2[1-e^2-g^2+(ab+e^2+g^2)x]}{\ell^2(b-1)}\,.
\end{align}
Upon a further rescaling of the $t$ and $\phi$ coordinates, we obtain the charged rotating black bottle solution (\ref{metric_charged_droplet}). The other solutions (\ref{metric_rotating_droplet}) and (\ref{metric_static_droplet}) can be obtained as special cases of (\ref{metric_charged_droplet}).

\bigskip\bigskip

{\renewcommand{\Large}{\normalsize}
}

\end{document}